\documentclass[sigconf]{acmart}
\AtBeginDocument{%
  }
\usepackage{makecell}

\copyrightyear{2025}
\acmYear{2025}
\setcopyright{cc}
\setcctype{by}
\acmConference[SC '25]{The International Conference for High Performance
Computing, Networking, Storage and Analysis}{November 16--21, 2025}{St
Louis, MO, USA}
\acmBooktitle{The International Conference for High Performance Computing,
Networking, Storage and Analysis (SC '25), November 16--21, 2025, St Louis,
MO, USA}
\acmDOI{10.1145/3712285.3759801}
\acmISBN{979-8-4007-1466-5/2025/11}

\acmSubmissionID{34}

\begin{document}

\title{DRIM-ANN: An Approximate Nearest Neighbor Search Engine based on Commercial DRAM-PIMs}


\author{Mingkai Chen, Tianhua Han, Cheng Liu \textsuperscript{*}, Shengwen Liang, Kuai Yu, Lei Dai, Ziming Yuan, Ying Wang, Lei Zhang, Huawei Li, and Xiaowei Li}
\thanks{*Corresponding author}
\affiliation{%
  \institution{SKLP, Institute of Computing Technology, Chinese Academy of Sciences}
  \city{Beijing}
  \country{China}}
\affiliation{%
  \institution{University of Chinese Academy of Sciences}
  \city{Beijing}
  \country{China}}
\email{{chenmingkai20s, hantianhua22s, liucheng, liangshengwen, dailei19z, yuanziming22s, wangying2009, zlei, lihuawei, lxw}@ict.ac.cn}
\email{kuaiyunee@outlook.com}

%
\renewcommand{\shortauthors}{Chen et al.}

\begin{abstract}
  Approximate nearest neighbor search (ANNS) is essential for applications like recommendation systems and retrieval-augmented generation (RAG) but is highly I/O-intensive and memory-demanding. CPUs face I/O bottlenecks, while GPUs are constrained by limited memory. DRAM-based Processing-in-Memory (DRAM-PIM) offers a promising alternative by providing high bandwidth, large memory capacity, and near-data computation. This work introduces DRIM-ANN, the first optimized ANNS engine leveraging UPMEM’s DRAM-PIM. While UPMEM scales memory bandwidth and capacity, it suffers from low computing power because of the limited processor embedded in each DRAM bank. To address this, we systematically optimize ANNS approximation configurations and replace expensive squaring operations with lookup tables to align the computing requirements with UPMEM’s architecture. Additionally, we propose load-balancing and I/O optimization strategies to maximize parallel processing efficiency. Experimental results show that DRIM-ANN achieves a 2.46× speedup over a 32-thread CPU and up to 2.67× over a GPU when deployed on computationally enhanced PIM platforms.
\end{abstract}



\keywords{Processing in memory (PIM), DRAM PIM, ANNS, approximate computing}


\maketitle

\section{Introduction}

Approximate nearest neighbor search (ANNS) enables effective similarity search on diverse data modalities, including images, graphs, and text, and has proven effective in critical applications such as recommendation systems and retrieval-augmented generation (RAG) in large language model (LLM)-based systems. Typically, it has the data of diverse modalities embedded with high-dimensional vectors and leverages the Euclidean distance between the vectors to characterize the data correlation. Thereby, ANNS involves searching massive vector corpora, leading to high computational complexity. This makes the processing both I/O-intensive and memory-demanding, necessitating significant algorithmic and system-level optimizations \cite{ref8}.

Extensive research has been conducted on optimizing ANNS for CPUs \cite{ref3, ref4, ref19, ref22} and GPUs \cite{ref4, ref5, ref16, ref17, ref18}. Notably, Faiss-CPU \cite{ref4} utilizes CPU vectorization and multi-threading processing to enable parallel ANNS computation, providing an industrial-grade high-performance ANNS library. JUNO \cite{ref5} exploits both RT cores and CUDA cores for lookup table (LUT) construction and distance calculation, respectively, to improve ANNS efficiency. However, CPU-based ANNS is often constrained by memory I/O bandwidth around a few dozens of GB/s. GPUs offer significantly higher bandwidth up to TB/s, but are typically limited by memory capacity and fail to accommodate large high-dimensional vectors required in many practical ANNS scenarios \cite{ref16, ref17, ref18}. In addition, customized ANNS accelerators based on FPGAs \cite{ref7, ref8, ref20, ref21} have also been proposed to improve computational efficiency. Despite notable performance gains, they remain constrained by either I/O bandwidth or memory capacity.

In this context, processing-in-memory (PIM) architectures—parti-cularly DRAM-based PIMs, which provide high bandwidth, large capacity, and near-data computing capabilities—have emerged as promising solutions for ANNS. Recent works such as GCiM \cite{ref20} and Pyramid \cite{ref22} have explored PIM for ANNS by integrating specialized computing fabrics with high-bandwidth memory, demonstrating substantial performance and energy efficiency improvements. However, these customized PIM architectures still face significant technological challenges and are primarily evaluated through simulators, which may sometimes yield overly optimistic results.

Inspired by prior PIM-based ANNS optimizations, we explored the feasibility of leveraging off-the-shelf DRAM-PIM from UPMEM for ANNS acceleration. A typical UPMEM system consists of 2,560 data processing units (DPUs), each equipped with 64 MB of memory and 1 GB/s bandwidth, resulting in a total memory capacity of 160 GB and an aggregate memory bandwidth of 2.56 TB/s \cite{ref11}. This configuration can scale linearly if the host machine supports additional DIMM slots, offering high bandwidth, large capacity, and near-data computing—key attributes for efficient ANNS execution. Unlike custom PIM architectures tailored for specific applications, UPMEM integrates a sequential RISC processor within each DPU, enabling general-purpose computing. However, these processors operate at only 450 MHz, achieving a nominal performance of approximately one instruction per cycle, despite supporting multi-threading \cite{ref11}. Performance degrades significantly for multiplication and division due to the lack of native hardware support. For instance, multiplication on UPMEM is approximately 32 times more expensive than addition.

The computing of each UPMEM DPU is inherently limited to the attached RISC-V processor, resulting in a low compute-to-I/O ratio across the overall PIM architecture. Consequently, while ANNS is typically I/O-intensive on CPUs, it can still be constrained by computational limitations on UPMEM. Despite the thousands of parallel DPUs, each with private memory enabling parallel execution, UPMEM lacks a flexible communication fabric between DPUs, unlike many customized PIM architectures. Additionally, DPUs often require synchronization before communicating with the host processor, as host-DPU interactions are costly, with bandwidth limited to just 0.75\% of the total PIM bandwidth. As a result, DPU-to-DPU communication is expensive, making load balancing crucial to prevent underutilization. This challenge is particularly pronounced in ANNS, where each query may access only a small subset of the dataset, potentially engaging only a few DPUs and leading to severe load imbalance. Furthermore, each DPU has a private memory hierarchy, necessitating careful data management to fully exploit the benefits of the memory hierarchy. In summary, optimizing ANNS for UPMEM presents significant challenges that require careful consideration of computational limitations, memory management, and load balancing.

\begin{table}[!t]
  \centering
  \caption{Large-scale ANNS datasets.}
  \label{table3}
  \begin{tabular}{l@{\ \ \ }c@{\ \ \ }c@{\ \ \ }c@{\ \ \ }c@{\ \ \ }c@{\ \ \ }c}
    \hline
    \textbf{Dataset} & \textbf{ST1B} & \textbf{DP1B} & \textbf{SV1B} & \textbf{T2I1B} & \textbf{ST100M} & \textbf{DP100M}\\
    \hline
    Vectors & $10^9$ & $10^9$ & $10^9$ & $10^9$ & $10^8$ & $10^8$ \\
    Dim & 128 & 96 & 100 & 200 & 128 & 96 \\
    \hline
  \end{tabular}
\end{table}

In this work, we propose DRIM-ANN, the first off-the-shelf DRAM-PIM-based framework for ANNS tasks. We begin by analyzing the computational patterns of ANNS with cluster-based index and designing an ANNS engine that supports Inverted File index with Product Quantization (IVF-PQ) and its variants, including OPQ \cite{ref48} and DPQ \cite{ref49}. To leverage UPMEM’s high memory bandwidth while mitigating its limited computational power, we replace the expensive Euclidean distance calculations typically required in ANNS by lossless square lookup tables (SQTs), significantly reducing computational overhead. On top of this optimized ANNS engine, we introduce a PIM-aware algorithm tuning framework that optimizes ANNS configurations to the target DRAM-PIM architecture. Specifically, we develop an effective ANNS performance model to guide architecture-aware tuning. Additionally, we address the substantial load balancing challenges associated with deploying ANNS on UPMEM, which features over 2,000 parallel DPUs. Basically, we propose cluster splitting and cluster duplication strategies that optimize data layout and distribute active computations more evenly across DPUs, alleviating load imbalance. Notably, we formulate the layout optimization problem and introduce a mixed-layout strategy for improved efficiency. Furthermore, we prioritize the duplication of frequently accessed clusters and dynamically allocate queries at runtime to balance the computational workload across DPUs more effectively.

In summary, we make the following major contributions.
\begin{itemize}
\item We present DRIM-ANN, the first ANNS framework leveraging commodity DRAM-PIMs. The proposed framework supports general cluster-based ANNS variants utilizing product quantization (PQ) and its extensions, such as OPQ \cite{ref48} and DPQ \cite{ref49}.

\item We propose a PIM-aware ANNS algorithm tuning strategy in combination with multiplier-less conversion based on fine-grained modeling of ANNS performance to lower the computing requirements of ANNS and alleviate the computing bottleneck of DRAM-PIMs.

\item We identify the load imbalance issue in ANNS implementations on DRAM-PIMs and introduce a systematic data layout optimization strategy combined with a runtime scheduling approach. These techniques mitigate query-induced imbalance and enhance the utilization of massively parallel DPUs in DRAM-PIMs.

\item Experimental results on representative datasets show that DRIM-ANN achieves up to $2.46\times$ speedup over a 32-thread CPU implementation. When scaled to SK Hynix's AiM \cite{ref53}, DRIM-ANN achieves a speedup of $2.67\times$ over an NVIDIA A100 GPU, demonstrating its potential for further acceleration on future DRAM-PIM architectures.

\end{itemize}

\section{Motivation}
\label{sec:motivation}
\subsection{Challenges of Typical ANNS on CPUs and GPUs}
\label{sec:intro-IVFPQ}

\begin{figure*}[!t]
\centering
\includegraphics[width=6.5in]{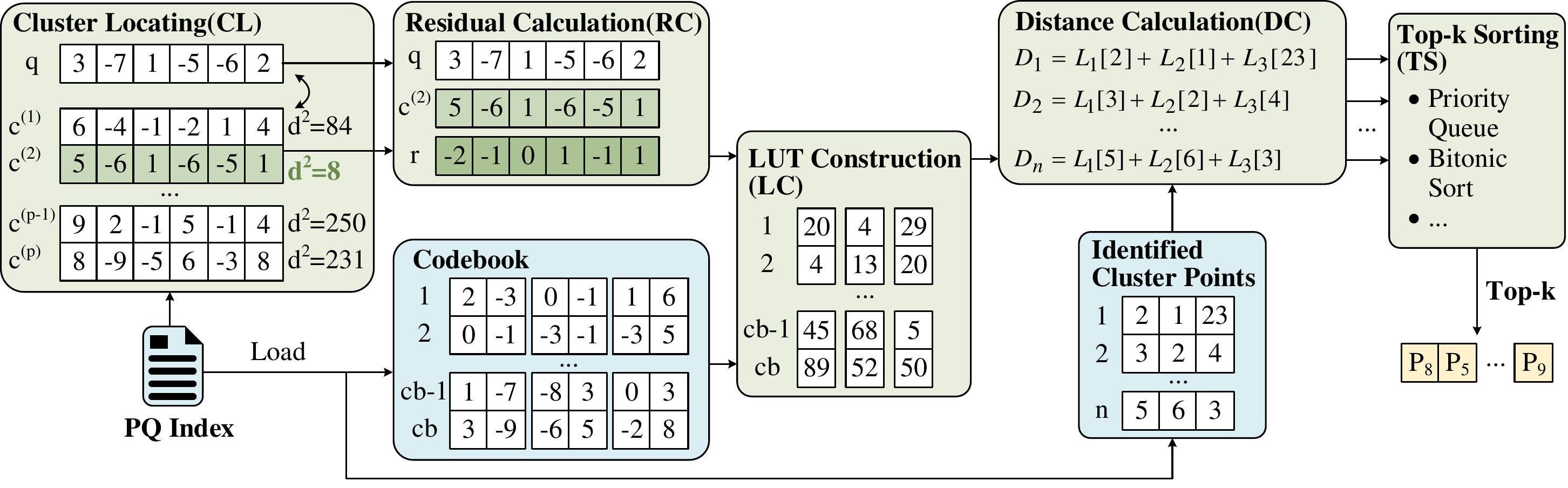}
\caption{Illustration of ANNS with cluster-based PQ index (L2 distance). Suppose there are $p$ clusters, The query is notated as q, and the centroid of the $i$ th cluster is notated as $c^{(i)}$. The L1 distance and the residual between q and $c^{(i)}$ are notated as d and r respectively. The amount of codebook entries and identified vectors to be scanned are notated as cb and n respectively.}
\label{fig2}
\end{figure*}

To efficiently perform ANNS on large datasets, cluster-based index is widely utilized, which is considered as the most efficient for billion-scale datasets with the same memory footprint \cite{ref1, ref50}. This kind of index typically divides the vector corpus into clusters and searches exhaustively within the clusters closest to the query. Since the number of data in the selected clusters is often large, product quantization (PQ) is applied to achieve both efficient storage and fast distance computation, which partitions high-dimensional vectors into smaller sub-vectors, clusters each subspace to construct a codebook, and encodes original vectors as index combinations of codebook entries across subspaces. 

The general processing of ANNS with cluster-based PQ index is illustrated in Figure~\ref{fig2}. It mainly involves five phases: cluster locating (CL), residual calculation (RC), lookup table (LUT) construction (LC), distance calculation (DC), and top-k sorting (TS). In CL, queries are assigned to several nearest clusters based on the L2 distance between the query and cluster centroids. In RC, the residual between the query and centroid is computed. Then, LC starts, where the L2 distance between the residual vector and a codebook in each subspace is calculated to form a distance LUT. Specifically, in each of the $M$ subspaces, the distance between the corresponding subvector of the residual and each of the $cb$ codebook subvectors is calculated, forming $cb$ LUT elements. In the example shown in Figure~\ref{fig2}, $M = 3$. DC computes the distance between the query and each cluster point by accumulating partial distances stored in the LUT that are identified by the corresponding encoded cluster point. Finally, in TS, the results are ranked, and the top-k nearest neighbors are obtained.

\begin{figure}[!t]
\centering
\includegraphics[width=3.3in]{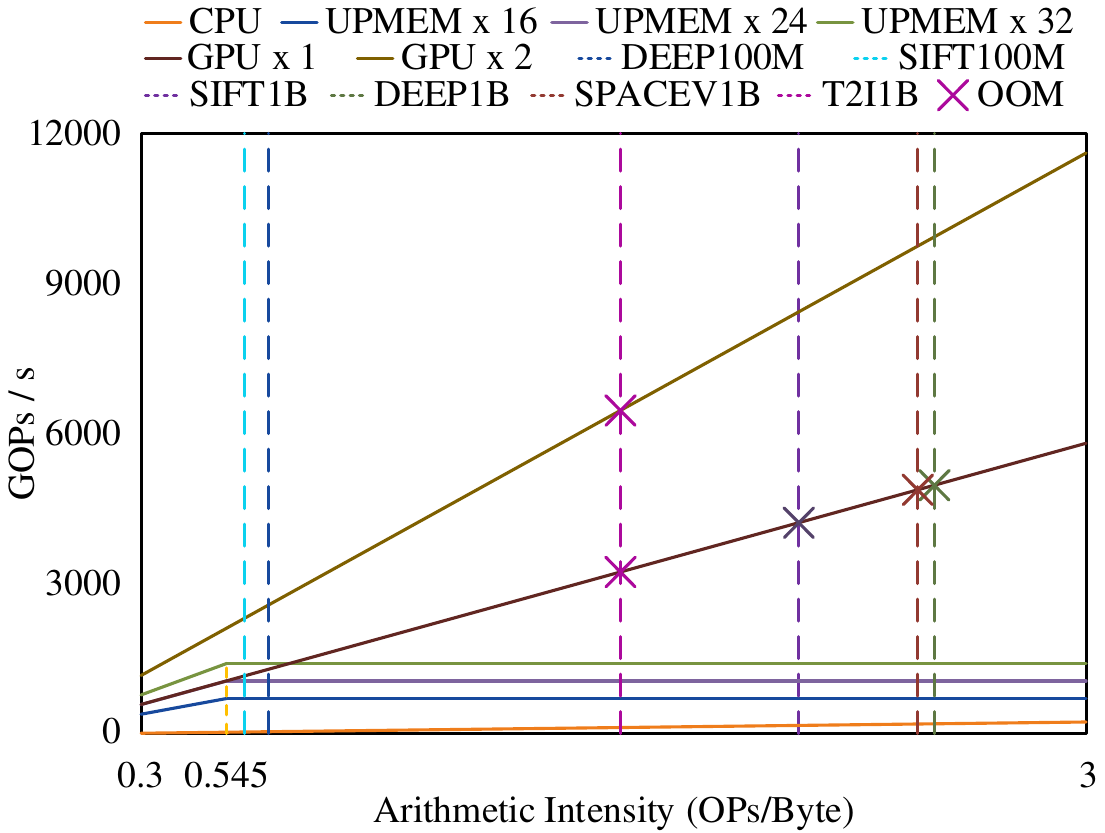}
\caption{The roofline analysis of ANNS on various platforms with several datasets. Intersection points correspond to the roofline performance whereas cross markers "x" indicate out of memory (OOM) errors per dataset-platform combination.}
\label{fig1}
\end{figure}

To identify the bottleneck of this processing, we conducted a roofline analysis of ANNS with cluster-based PQ index based on the Faiss library \cite{ref4} on several datasets \cite{ref65} listed in Table~\ref{table3}. As shown in Figure~\ref{fig1}, CPU is bounded by memory accesses across all datasets, limiting its performance compared to other platforms. Although GPU performs well on SIFT100M and DEEP100M, it suffers out-of-memory (OOM) error on larger datasets. Even though it is possible to save the data in memory on CPU side, it will induce frequent GPU-CPU data transfers and suffers PCI-E bandwidth limit accordingly. Meanwhile, it results in a significant waste of computational power and memory bandwidth to scale GPU with additional cards for capacity, whereas UPMEM achieves proportional increases in computational power, memory bandwidth and capacity by installing several additional DIMMs on the motherboard. As shown in Figure~\ref{fig1}, UPMEM achieves comparable bandwidth to an NVIDIA A100 GPU through 24 DIMMs, and it can scale computational power with 8 DIMMs added to surpass a GPU on SIFT100M and DEEP100M. The adaptive scalability‌ of UPMEM seamlessly aligns with ANNS's diverse demands of computational ability, memory bandwidth and capacity under various scenarios, exhibiting significant potential in acceleration through resource reconfiguration. Nevertheless, as indicated in Figure~\ref{fig1}, ANNS is constrained by the computational power of UPMEM, necessitating a reduction in computational loads to enhance throughput.

\begin{figure}[!t]
\centering
\includegraphics[width=3.3in]{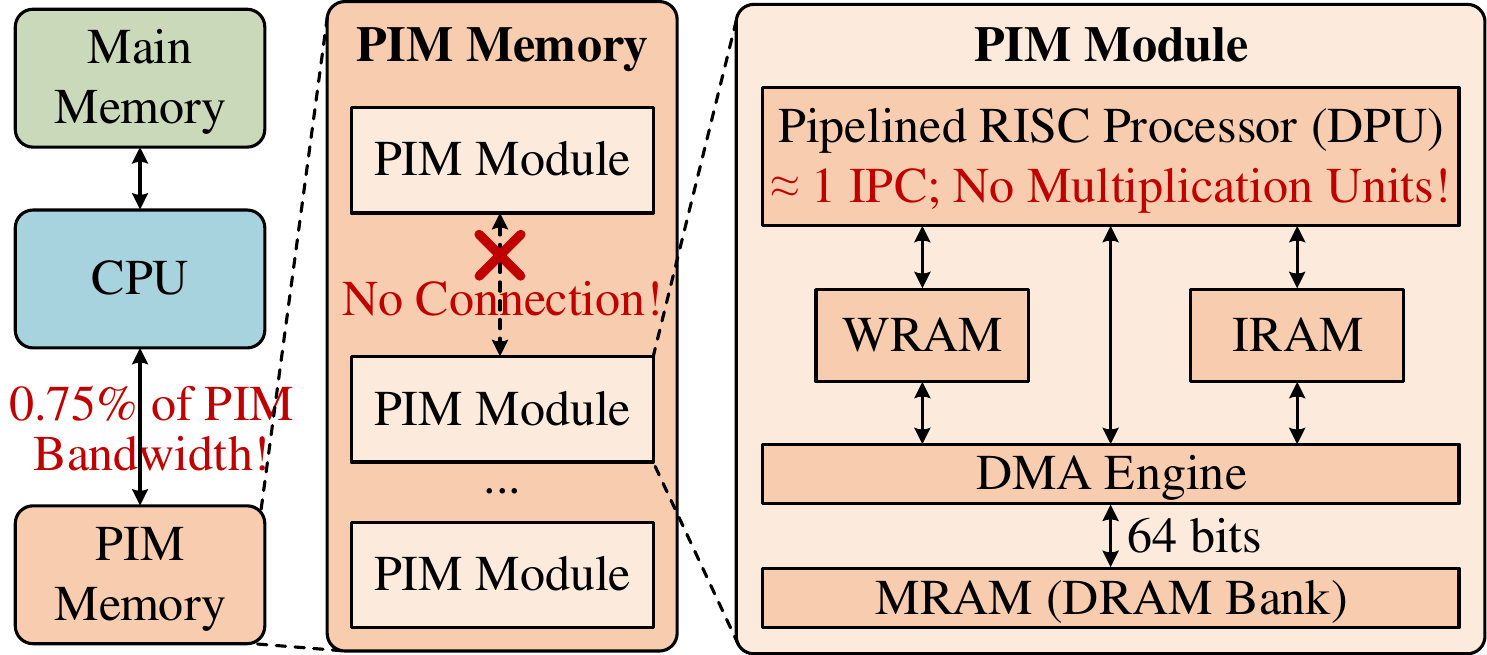}
\caption{UPMEM architecture.}
\label{fig6}
\end{figure}

\subsection{Hardware Characteristics of DRAM-PIMs}
\label{sec:intro-PIMs}
\label{sec:affinity-analysis}

DRAM-PIMs enhance memory bandwidth by embedding processing units into memory chips to minimize wirelength between compute units and memory cells. Specifically, UPMEM has introduced the first DDR4-PIM product, whose combined memory bandwidth achieves 2.56TB/s with 2,560 in-memory DPUs on a typical system \cite{ref10}. 64KB working memory (WRAM) is allocated as a scratchpad buffer to every 64MB DRAM memory (MRAM) on UPMEM chips for bandwidth amplification through data locality. The architecture of UPMEM is illustrated in Figure~\ref{fig6}. 

Despite the high bandwidth provided by PIM products, they are confronted with limited computational capabilities and high transfer overheads. Firstly, the nominal performance of each DPU is only about 1 instruction per cycle \cite{ref11} due to the limit of on-chip area and power, although each DPU supports tens of threads. Thus, the vector operations constituting the majority of ANNS require serial dimensional processing. Worse yet, lack of dedicated multiplication units on DPUs imposes 32 cycles per multiplication on distance calculation. As for data transfer between the host and DPUs, the bandwidth is limited by both memory channels and I/O bandwidth of host DRAM chips, which is no more than 0.75\% of the bandwidth inner PIM chips \cite{ref14}, limiting real-time data transfer and synchronization. Thus, load balance constitutes a critical determinant for sufficient utilization of combined PIM units, which may be violated by the imbalance of cluster-based indices and query distribution in ANNS. Although Samsung and SK-Hynix proposed HBM-PIM solutions \cite{ref52, ref53} with better computational ability, the gap between them and advanced GPUs is still huge, and they also suffer from limited HBM capacity and expensive host-HBM interconnection. These challenges motivate our systematic exploration of the inherent bandwidth advantages of DRAM-PIMs for ANNS in this work.

\section{DRIM-ANN Framework}
\label{sec:DRIM-ANN-LB}
\begin{figure}[!t]
\centering
\includegraphics[width=3.3in]{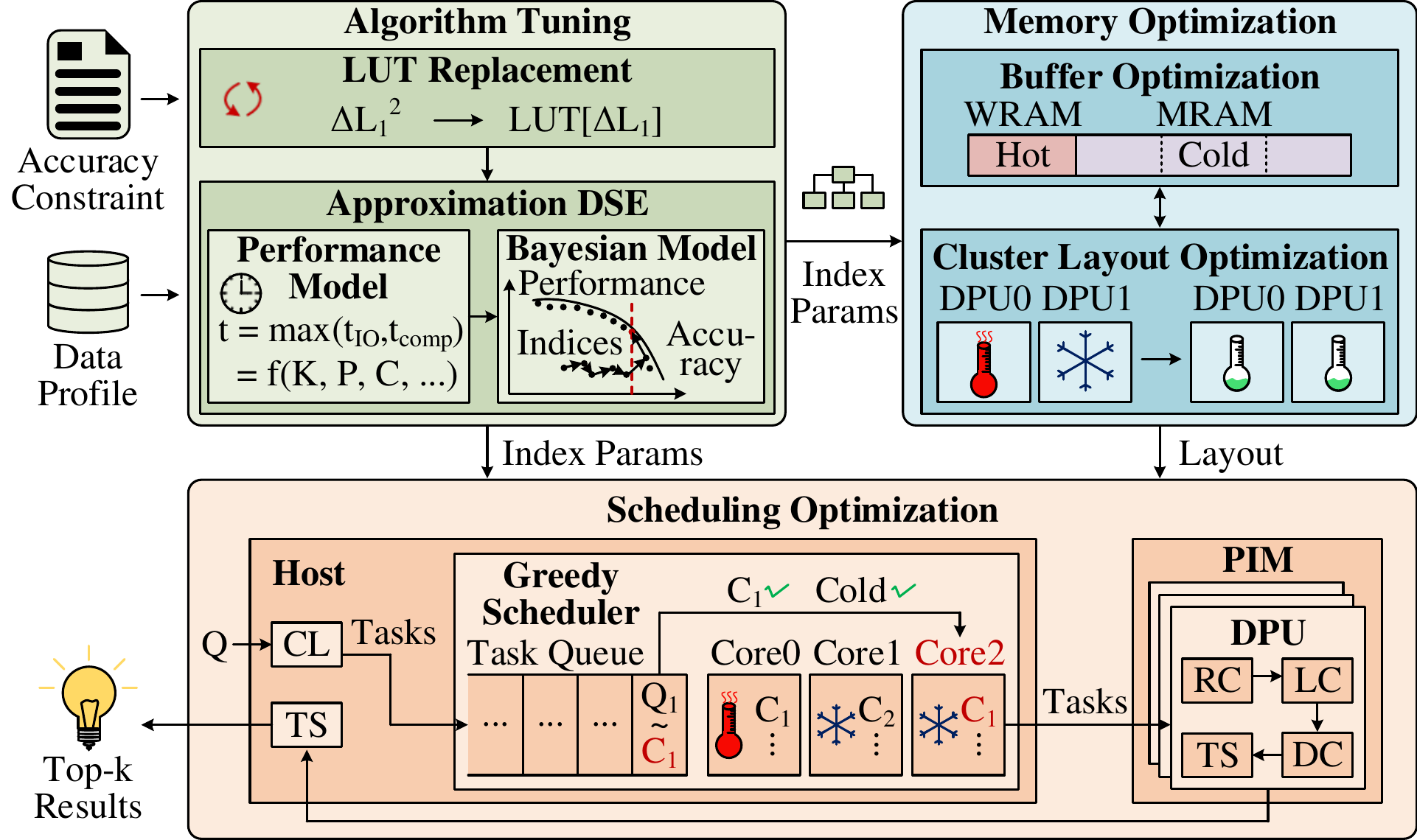}
\caption{The overview of DRIM-ANN framework.}
\label{fig3}
\end{figure}

To exploit UPMEM's high memory bandwidth and capacity that are both critical for ANNS, we propose DRIM-ANN, the first ANNS framework built on commodity DRAM-PIM architecture from UPMEM, as illustrated in Figure~\ref{fig3}. DRIM-ANN first introduces algorithmic optimizations to reduce computational demands, thereby aligning with the limited processing capabilities of UPMEM for higher performance. It then leverages the access frequency variation across the massive vectors inherent in each ANNS query and places hot vectors in WRAM with lower access latency to suit the DPU memory hierarchy and enhance the memory access efficiency. In parallel, it refines data layout across DPUs based on the data access frequency with both duplication and splitting to make sure the DPUs are evenly utilized as much as possible. Finally, DRIM-ANN integrates runtime query scheduling with data layout optimizations to mitigate skewed processing and address load imbalance caused by uneven query distribution across DPUs.

\subsection{Algorithm Tuning}
To address the computational limitations of UPMEM, we tune ANNS to reduce its computational intensity and adjust the computing requirements to better suit UPMEM from two complementary perspectives. First, we observe that distance calculation, which is a key operation in both the cluster locating and cluster searching phases, is computationally expensive on UPMEM due to the lack of native multipliers. Notably, the multiplications involved in L2 distance calculations are essentially squaring operations. To mitigate this overhead, we replace them with a squaring lookup table (SQT). When ANNS vectors are quantized to 8-bit integers, the SQT contains only 256 entries, which can be entirely stored in the on-chip memory for efficient access. For 16-bit quantization, where the SQT exceeds WRAM capacity, we store the most frequently used entries in WRAM and the remainder in MRAM. Since the squaring operands are the residuals between vectors, their values typically fall within a narrow range, allowing the WRAM-resident portion of the LUT to handle most cases effectively. Overall, the LUT-based replacement significantly reduces computational demands, at the cost of a modest increase in memory bandwidth consumption.

Beyond the LUT optimization, we leverage the inherent approximative nature of ANNS, which offers a broad design space including variations in vector dimension, data width, and product quantization. Based on this observation, we perform a design space exploration to identify configurations that reduce computational requirements while preserving search accuracy. This exploration further enhances ANNS performance by tailoring it to the architectural characteristics of UPMEM. The details of this approximation design space exploration are presented in the following section.

\begin{figure*}[!t]
\centering
\includegraphics[width=7in]{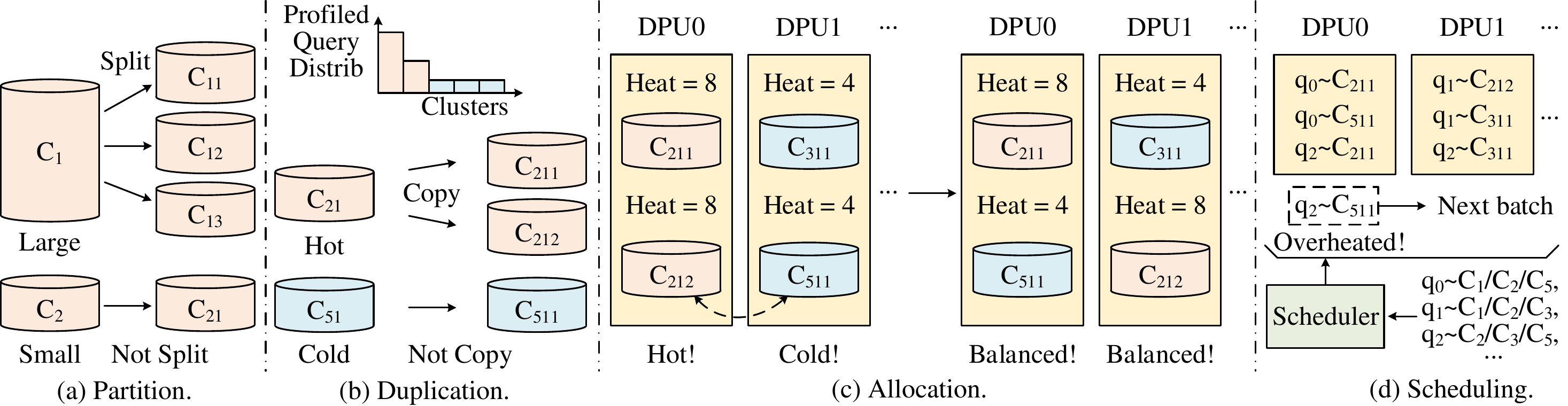}
\caption{An example of the proposed load-balance strategy.}
\label{fig5}
\end{figure*}

\subsection{Memory Optimizations}
Unlike CPUs and GPUs that have fully interleaved memory accessible to applications, the memory in UPMEM is non-interleaved and memory optimizations are critical to make full use of the theoretical bandwidth. While ANNS generally has the vectors clustered and each cluster is stored sequentially to ensure better access locality, the cluster access frequency may vary substantially because some of the clusters can be hot in many practical application scenarios and are more frequently accessed accordingly. Hence, the clusters located in different DPUs may exhibit distinct access frequency, leading to both low memory bandwidth utilization and DPU utilization. 

In order to address this problem, we investigate the data layout optimization based on the cluster access frequency obtained from profiling. Specifically, each cluster is split into several slices with equal capacity at first, since the size of different clusters may vary greatly. Although the size of each slice may still be different, the difference becomes smaller than the capacity of cluster slices, alleviating the imbalance among cluster points. Then, each cluster is duplicated several times and each copy is assigned to different DPUs. The duplicate factor for each cluster is determined by the heat profiled by random data distribution patterns so that hot clusters have more copies than cold ones. Finally, each cluster slice is allocated to the coldest DPU in order to balance the heat of each DPU. At runtime, the processing latency of retrieving the neighbors of a query in each of its closest clusters is predicted with a performance model, and the query is assigned to the coldest DPU where a copy of the corresponding cluster is located to balance the workloads among DPUs. An example is shown in Figure~\ref{fig5}.

\textbf{Cluster partition.} In this phase, large clusters are split into small slices to alleviate the load imbalance caused by point distribution. We set the maximum permitted size of these cluster slices as a threshold $th_1$. Figure~\ref{fig5}(a) presents an example of cluster partition. The size of the cluster $C_1$ is larger than $th_1$, so $C_1$ is split into three slices i.e. $C_{11}$, $C_{12}$ and $C_{13}$, whose sizes are no larger than $th_1$. The size of $C_2$ is smaller than $th_1$, so $C_2$ is not split. With cluster partition, the workloads on $C_1$ can be assigned to a maximum of three DPUs instead of only one DPU, and the workloads on each DPU are approximately the same as those where $C_2$ locates, leading to more flexible and balanced assignment of online workloads on $C_1$. The splitting threshold $th_1$ is obtained by iterations to trade off the extra indexing overheads with the benefits from cluster partition under the constraint that all of the metadata are able to be placed on WRAMs. Specifically, $th_1$ is set as the size of the smallest cluster at the beginning and iterates with a dynamic learning rate. The extra indexing overheads of the metadata are estimated by the product of the access times of each slice and the latency of each access, where the access times of each cluster are estimated by profiling with random data distribution and the latency of each access is profiled offline.

\textbf{Cluster duplication.} In this phase, hot clusters are duplicated several times to alleviate the load imbalance caused by query distribution. The heat of each cluster is estimated by the weighted sum of its size and its heat profiled with random data distribution. As shown in Figure~\ref{fig5}(b), the hot cluster $C_2$ is duplicated and has one copy, while the cold cluster $C_5$ has no copies. If a hot cluster is split into several slices, every slice should be duplicated the same times. The duplicated times $th_2[i]$ of the $i$ th cluster is proportional to its heat and it is in inverse proportion to its amount of split slices, since for each cluster, the goal of cluster duplication is to assign each pair of its slices and queries on it to distinct DPUs at runtime. To provide more choices for runtime scheduling, we try to produce as many duplicated cluster slices as PIM memory allows in this phase.

\textbf{Cluster allocation.} In this phase, cluster slices and their copies are allocated to DPUs to balance the heat among DPUs. The heat values of the cluster slices on a DPU are accumulated to form the heat of the DPU. For example, in Figure~\ref{fig5}(c), $C_{211}$ and $C_{511}$ locate on $DPU_0$, so the heat of $DPU_0$ is 12 which is the sum of their heat. For the reuse of common data for each cluster such as the residual result, the distance look-up table and the priority queue which are used in RC, LC and TS respectively,  we try to allocate slices of the same cluster to the same DPU. Specifically, cluster slices are allocated to each DPU in order while keeping the balance of the heat of DPUs at the beginning. Then, we look up slices of the same cluster on different DPUs based on a mapping table and try to exchange them with other slices to locate them on the same DPU. Since the heat of the DPUs that they locate on may be reduced if the exchange succeeds, we balance the heat of all the DPUs again and start the next iteration until there are little chances for data reuse while keeping load balance. Finally, we allocate cluster slices to DPUs based on the optimized mapping table.

\begin{figure*}[!t]
\centering
\includegraphics[width=7in]{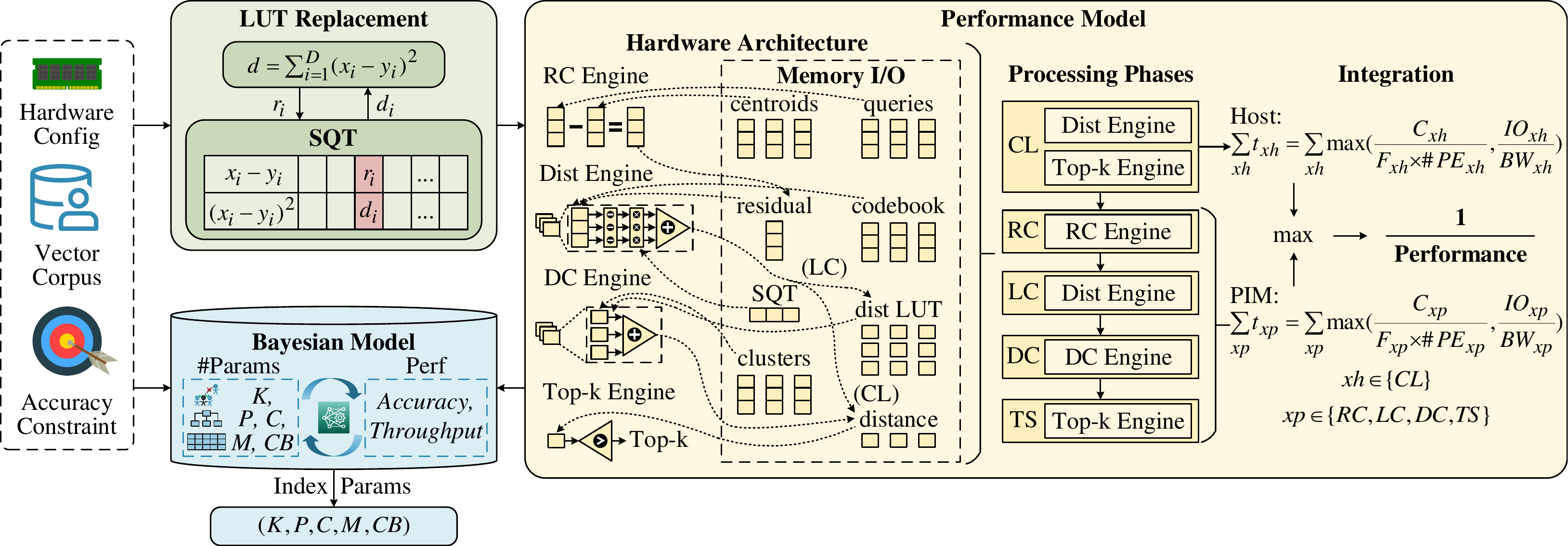}
\caption{The Proposed ANNS Tuning Framework. It optimizes ANNS for UPMEM by reducing its computing-to-I/O ratio through LUT replacement and approximation design space exploration. }
\label{fig20}
\end{figure*}

In order to improve the performance of ANNS, we also investigate the memory operations of each DPU. 
To make the real performance approach the ideal performance of the optimal index, we optimize the utilization of WRAM buffer and parallel pipeline of each DPU for improvement of the real bandwidth. As the capacity of WRAM buffer is only 0.1\% of PIM memory, only a few data can be placed on it. To make the best use of it, we estimate the access times of each kind of data such as the codebook and centroids by the coefficient of I/O in Equation~\ref{eq1}-\ref{eq10}. The heat of each kind of data is represented as the average access times per bit, and the hottest data are placed on WRAM buffer. Furthermore, we assign workloads over different codebook entries or cluster points to multiple threads to fulfill the pipelines of each DPU while minimizing extra overheads of synchronization and WRAM buffer, as the distance engine and DC engine illustrate in Figure~\ref{fig20}. 

\subsection{Scheduling Optimizations}
At runtime, DRIM-ANN equips a greedy scheduler to map queries to DPUs. Each query is mapped to the coldest DPU where the closest clusters to it are located. The heat of DPUs is estimated by the latency calculated by Equation~\ref{eq1}-\ref{eq11} where $Q, P, C$ are replaced by the real-time values on each DPU. After greedy scheduling, the scheduler tries to schedule imbalanced workloads to other DPUs with duplicated slices to lower the long-tail latency. Then, the tasks that make the heat of DPUs exceed a threshold $th_3$ over the average heat will be postponed to the next batch. As depicted in Figure~\ref{fig5}(d), $DPU_0$ is overheated, so the task of searching k neighbors of $q_2$ in the cluster slice $C_{511}$ is postponed to the next batch. After query scheduling, all DPUs are triggered synchronously to execute the allocated tasks.

\section{Approximation Design Space Exploration}
\label{sec:algorithm-tuning}

After multiplier-less conversion on DRAM-PIMs, several algorithm parameters are tunable to balance performance and accuracy of ANNS, and the effects of each parameter can be compensated for by other parameters. Table~\ref{table1} lists these parameters. Among the notations, $K, P, C, M, CB$ are tunable parameters of the cluster-based index. Besides, $N, Q, D, B_{x}$ specify the shape of the dataset, and $BW_{x}, \#PE, F_{x}$ are hardware parameters related to the PIM platform, which are fixed for a specific dataset and PIM platform.

\begin{table}[!t]
  \centering
  \caption{Notations for DRIM-ANN framework.}
  \label{table1}
  \begin{tabular}{l@{\ \ }l@{}}
    \hline
    \textbf{Notation} & \textbf{Description}\\
    \hline
    $N$ & The amount of clusters on a PU \\
    $Q$ & The amount of queries on a PU \\
    $D$ & The dimension of queries and centroids \\
    $B_{x}$ & \makecell[l]{The bit width of $x$. $x \in \{c(=centroid), q(=query), $\\$p(=point), cb(=codebook), l(=LUT), a(=address)\}$}\\
    $K$ & The amount of nearest neighbors for each query \\
    $P$ & \makecell[l]{The amount of located clusters of each query on \\a DPU} \\
    $C$ & The average point amount of clusters \\
    $M$ & The amount of subvectors divided by each query \\
    $CB$ & The amount of codebook entries \\
    $BW_{x}$ & \makecell[l]{The bandwidth of the phase $x$. $x \in \{CL(=cluster $\\$locating), RC, LC, DC, TS\}$} \\
    $\#PE$ & The amount of host threads or PIM DPUs \\
    $F_{x}$ & \makecell[l]{The frequency of $x$. $x \in \{CL(=cluster\ locating), $\\$RC, LC, DC, TS\}$} \\
    $C_{x}$ & \makecell[l]{The computation of the phase $x$. $x \in \{CL, RC, LC, $\\$DC, TS\}$} \\
    $IO_{x}$ & \makecell[l]{The memory accesses of the phase $x$. $x \in \{CL, RC, $\\$LC, DC, TS\}$} \\
    \hline
  \end{tabular}
\end{table}

Since it is costly to evaluate the effects of all parameter combinations on real DRAM-PIM platforms, especially when the scale of datasets is large, we propose a model to estimate the performance for the exploration of the parameter space. It also provides a heuristic guide for data layout among different memory hierarchies of DRAM-PIMs and task assignment between the host and DRAM-PIM platforms. As illustrated in Figure~\ref{fig2} and Figure~\ref{fig20}, distance calculation between queries and cluster centroids, and top-k sorting for the distances are performed in CL. These operations costs $C_{CL}$ cycles, which is:
\begin{equation}
\label{eq1}
    C_{CL} = Q \times N / C \times (dist(D) + (\log P - 1))
\end{equation}
where the cost of distance calculation between X-dimension vectors is estimated as:
\begin{equation}
\label{eq15}
    dist(X) = X \times 3 - 1
\end{equation}
since the subtraction, multiplication and accumulation are executed by elements at each dimension in the vector operands. The centroids and queries are demanded by distance calculation, and a priority queue whose size is $log P + 1$ is saved after top-k sorting, so the cost of memory accesses in CL is:
\begin{equation}
\label{eq2}
\begin{split}
    IO_{CL}& =Q \times N / C \times ((B_c + B_q) \times D \\
    &+ (B_l + B_a) \times (\log P + 1))
\end{split}
\end{equation}
Similarly, the residuals between each query and its top-$P$ nearest cluster centroids are calculated in RC, where the dimension of operand are $D$, so the cost is:
\begin{equation}
\label{eq3}
    C_{RC} = Q \times P \times D
\end{equation}
and the memory overhead is:
\begin{equation}
\label{eq4}
    IO_{RC} = (B_c + B_q) \times Q \times P \times D
\end{equation}
In LC, the residual is split to $M$ subvectors, whose dimension is $D / M$. The distance between each subvector and codebook entries in the corresponding subspace is calculated to form a distance LUT, so the cost is:
\begin{equation}
\label{eq5}
    C_{LC} = Q \times P \times CB \times dist(M) \times \frac{D}{M}
\end{equation}
where the residual and codebook are loaded from the memory, whose overhead is:
\begin{equation}
\label{eq6}
    IO_{LC} = Q \times P \times CB \times ((B_{cb} + B_q) \times D + B_l \times M)
\end{equation}
The distance results stored in the LUT is retrieved based on the identified points in the top-$P$ nearest clusters of each query and accumulated to get the actual distance, so the cost is:
\begin{equation}
\label{eq7}
    C_{DC} = Q \times P \times C \times (M - 1)
\end{equation}
where the points and the LUT are loaded, whose overhead is:
\begin{equation}
\label{eq8}
    IO_{DC} = Q \times P \times C \times ((B_a + B_l) \times M + B_l)
\end{equation}
Finally, a priority queue of the top-k list is updated by the actual distance, whose overheads are:
\begin{equation}
\label{eq9}
    C_{TS} = Q \times P \times C \times (\log K - 1)
\end{equation}
\begin{equation}
\label{eq10}
    IO_{TS} = (B_l + B_a) \times Q \times P \times C \times (\log K + 1)
\end{equation}

Since computation and memory accesses are overlapped on both the host and PIM platforms, the time for each phase is estimated as:
\begin{equation}
\label{eq11}
\begin{split}
    &t_{x} = max(\frac{C_{x}}{F_{x} \times \#PE_{x}}, \frac{IO_{x}}{BW_{x}}), \\
    &x \in \{CL, RC, LC, DC, TS\}
\end{split}
\end{equation}

\noindent
the C2IO of each phase is:
\begin{equation}
\label{eq12}
    C2IO_{x} = \frac{C_{x}}{IO_{x}}, x \in \{CL, RC, LC, DC, TS\}
\end{equation}

Despite the fact that most operations become memory-intensive after multiplier-less conversion, those with higher C2IO can be placed on the host to overlap other operations. The time for data transfer between the host and DRAM-PIMs is omitted in Equation~\ref{eq1}-\ref{eq11}, which should be avoided due to the poor transfer bandwidth and can be negligible in our experiments. The effectiveness of the performance model is verified in Section~\ref{sec:evaluation:optimizations}.

\subsection{ANNS Design Space Exploration}
\label{sec:ANNS-DSE}

On a specific dataset and PIM-platform, $N, Q, D, B_{x}$ and $BW_{x}, \#PE, F_{x}$ are fixed. Thus, the task of the design space exploration (DSE) for DRIM-ANN is to find a combination of $K, P, C, M, CB$ that achieves the optimal performance under the given accuracy constraint. Suppose the mapping from each group of ($K, P, C, M, CB$) to the accuracy as $a$, since the overheads on the host and DRAM-PIMs are overlapped, the task of DSE is:
\begin{equation}
\label{eq13}
\begin{split}
    minimize\ max(&\sum_{xh} t_{xh}, \sum_{xp} t_{xp}), \\
    s.t.\ a(K, P, C, M, CB) &\ge accuracy\_constraint
\end{split}
\end{equation}

\noindent
in which
\begin{equation}
\label{eq14}
\begin{split}
    & xh, xp \in \{CL, RC, LC, DC, TS\}, \\
    & xh\ on\ the\ host, xp\ on\ DRAM-PIMs
\end{split}
\end{equation}

Though the target function in Equation~\ref{eq13} can be derived by Equation~\ref{eq1}-\ref{eq11}, the accuracy function $a$ is ambiguous. Thus, we utilize Bayesian optimization \cite{ref33} to explore the optimal index. Each index is represented by a group of ($K, P, C, M, CB$). 
The proposed performance model serves as the performance estimation part of the kernel function, and the accuracy estimation part is modeled by Mat\'ern kernel function. They are combined and input to the Gaussian process as the surrogate model to estimate the performance and accuracy of each index. They are also utilized by the acquisition function which is based on expected hypervolume improvement (EHVI) \cite{ref54} to guide the next iteration. The surrogate model is trained with several samples \cite{ref8}. During the exploration, we select a group of ($K, P, C, M, CB$) within the accuracy constraint through greedy search as the initial index and explore the implicit space from it with several iterations. The user-provided accuracy constraint is required to be satisfied during the optimization. When the design space is small, the DSE process is similar to exhaustive search.

\section{Evaluation}
\label{sec:evaluation}
\subsection{Experiment Setup}
Our main experiments are conducted on UPMEM PIM-DIMM \cite{ref10} platform, which is equipped with a maximum of 2543@350MHz DPUs and 159GB UPMEM memory. The host CPU is equipped with Intel Xeon Silver 4216@2.1GHz processor and 256GB DDR4. We compare DRIM-ANN with the state-of-the-art cluster-based ANNS implementations including Faiss-CPU and Faiss-GPU \cite{ref4} on CPU and GPU respectively. CPU platform is equipped with Intel Xeon Gold 5218@2.3GHz processor and 512GB DDR4. GPU platform is equipped with an NVIDIA A100 PCIe and 80GB HBM2e. The CPU baseline is optimized with AVX/AVX2 and openmp library. The energy consumption of the baseline and the UPMEM server is obtained from Intel RAPL domain \cite{ref15}. To avoid data overflow of the baseline on the GPU memory, we adopt the 100-million-vector size scale for both SIFT \cite{ref12} and DEEP \cite{ref13} datasets. DEEP100M is quantified to uint8 to keep in coincidence with SIFT100M. We apply the query sets of SIFT1B and DEEP1B as queries to SIFT100M and DEEP100M respectively. Both of the query sets have 10000 queries. The dimension of SIFT100M and DEEP100M is 128 and 96 respectively. All experiments are under the accuracy constraint that recall@10 $\ge$ 0.8 advised by prior works like ANNA\cite{ref7} and FANNS\cite{ref8}.

\begin{figure}[!t]
\centering
\includegraphics[width=3.3in]{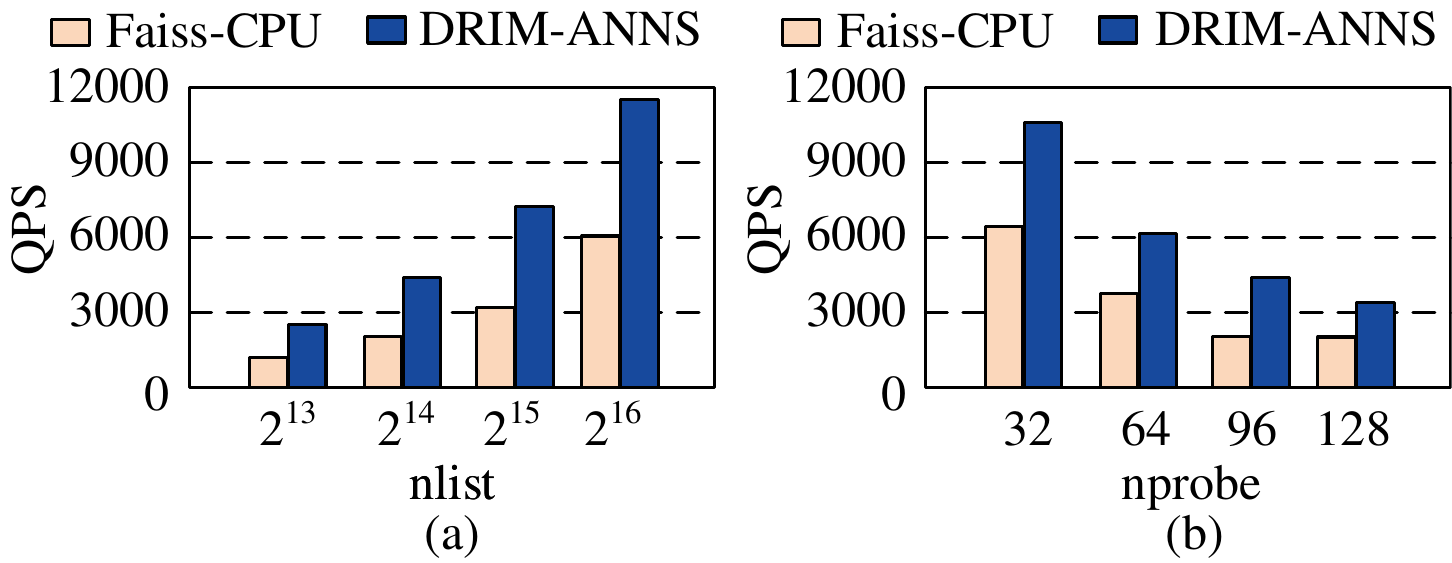}
\caption{End-to-end performance on SIFT100M.}
\label{fig11}
\end{figure}

\begin{figure}[!t]
\centering
\includegraphics[width=3.3in]{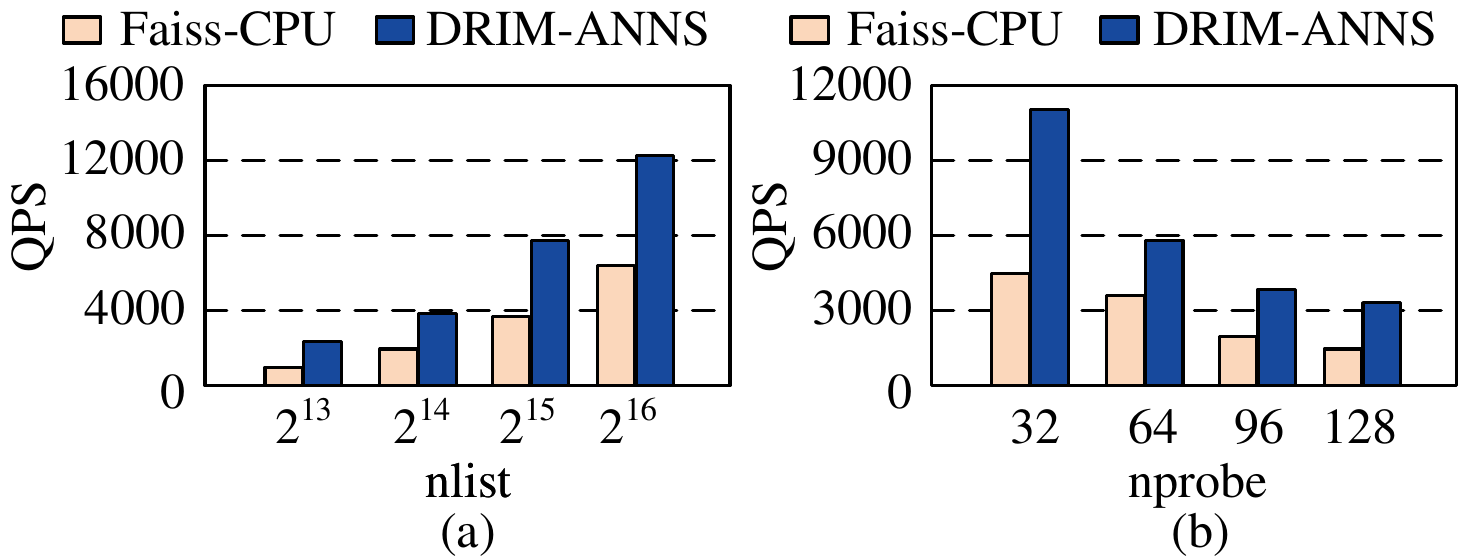}
\caption{End-to-end performance on DEEP100M.}
\label{fig12}
\end{figure}

\subsection{Performance Analysis}
\label{sec:evaluation:perfanalyse}
\textbf{End-to-end performance.} We first compare the performance of DRIM-ANN with Faiss-CPU and explore the effects of different algorithm parameters on the speedup. The amount of codebook entries is set as 256, which is required by Faiss-CPU, while DRIM-ANN supports more codebook entries. Each vector is split into 16 subvectors. Quantified DEEP100M is utilized by both DRIM-ANN and Faiss-CPU. Performance is measured in queries per second (QPS).

As shown in Figure~\ref{fig11} and Figure~\ref{fig12}, DRIM-ANN achieves 1.63X-2.25X and 1.61X-2.46X speedup over CPU on SIFT100M and DEEP-100M respectively. The geomean speedups are 1.89X and 2.08X on SIFT100M and DEEP100M respectively. In Figure~\ref{fig11}(a), each query retrieves 96 clusters, i.e. $nprobe$ = 96, and we set the amount of clusters $nlist$ = $2^{14}$ in Figure~\ref{fig11}(b). Since the size of AVX registers is fixed on CPU while the calculation on UPMEM is completed dimension by dimension, the computational ability on CPU is wasted on DEEP100M whose dimension is not a power of two, leading to a higher speedup of DRIM-ANN than that on SIFT100M. The speedup is higher with the increase of cluster size since more cluster points are retrieved in this case, which is especially friendly to larger datasets. The results indicate that DRIM-ANN brings considerable performance improvement with the use of DRAM-PIMs across different choices of cluster-based ANNS indices.

\begin{figure}[!t]
\centering
\includegraphics[width=3.3in]{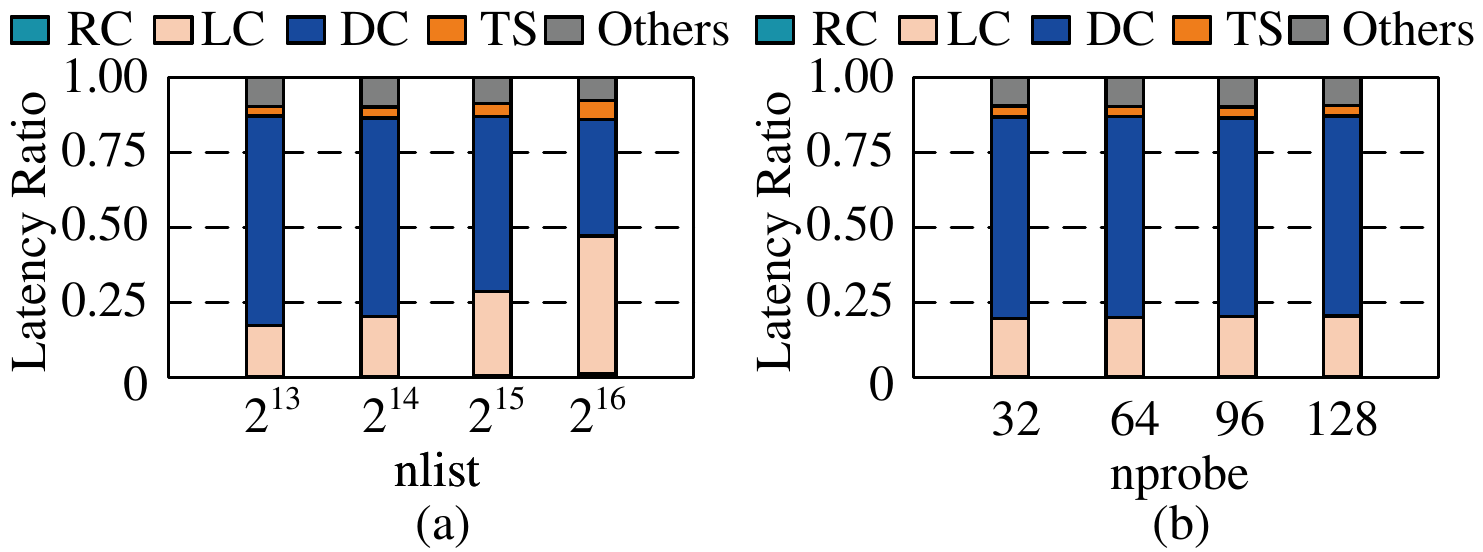}
\caption{Performance breakdown on SIFT100M.}
\label{fig13}
\end{figure}

\textbf{Performance analysis of PIM kernels.} To better understand the performance bottleneck of PIM kernels, we breakdown the PIM latency of DRIM-ANN at the granularity of kernels in Figure~\ref{fig13}. CL is implemented on the host and other four phases are placed on DPUs to balance the amount of transferred data and the utilization of both DPUs and the host CPU. The latency of host execution and data transfer between the host and DRAM-PIMs is fully overlapped with that of DPU execution among all the experiments. As shown in Figure~\ref{fig13}, LC and DC take the largest proportion of the overall latency, and the bottleneck changes from DC to LC with the increase of $nlist$. The reason is that both kernels are used frequently. As indicated by Equation~\ref{eq5}-\ref{eq8}, distance calculation of each dimension in LC takes place $Q \times P \times CB \times D$ times, while distance accumulation of each dimension in DC takes place $Q \times P \times C \times M$ times for each query batch. Since $C$ is in inverse proportion to $nlist$ on the same dataset, the increase of $nlist$ reduces the workloads in DC, leading to the change of bottleneck. The results indicate that optimization of LC and DC such as multiplier-less ANNS conversion and buffer optimization would bring considerable benefits.

\begin{figure}[!t]
\centering
\includegraphics[width=3.3in]{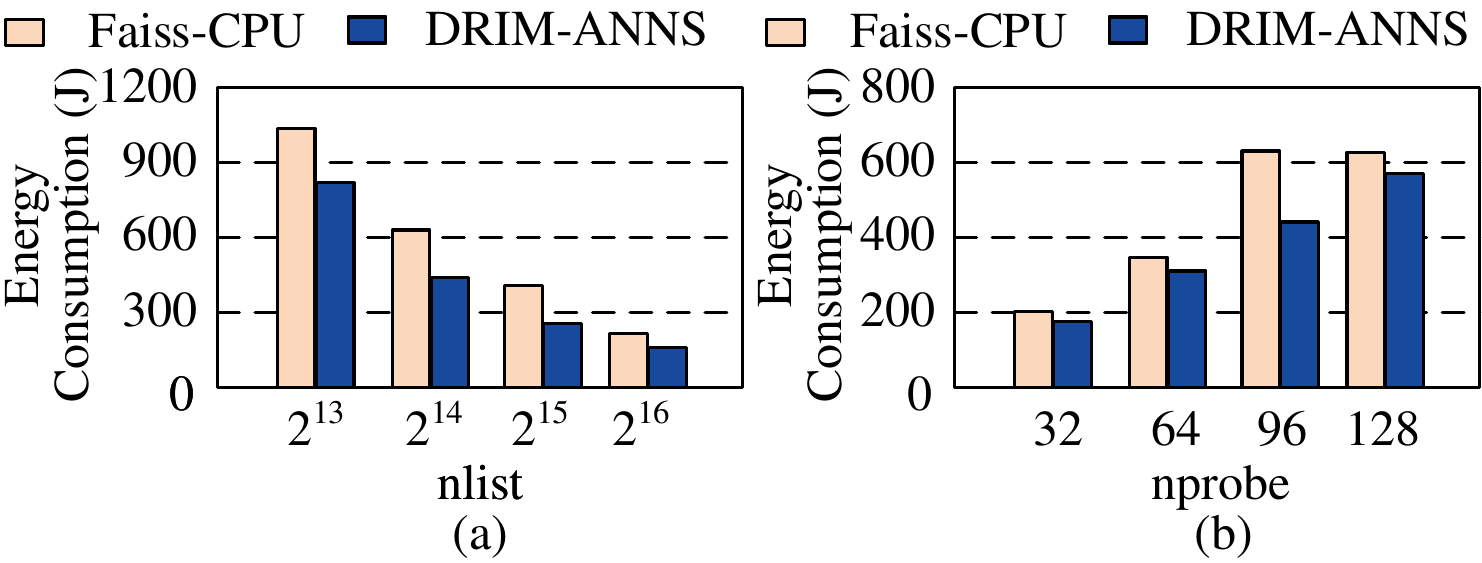}
\caption{End-to-end energy efficiency comparison.}
\label{fig19}
\end{figure}

\textbf{Energy efficiency.} We compare the energy consumption of DRIM-ANN against the CPU baseline on SIFT100M. As depicted in Figure~\ref{fig19}, DRIM-ANN achieves 1.10X-1.58X better energy efficiency over CPU and the geomean is 1.27X. Although each PIM-DIMM consumes about 13.92W \cite{ref11} so that the power of UPMEM server is higher than the CPU server, DRIM-ANN still achieves better energy efficiency than the CPU baseline due to better performance. On DEEP100M, DRIM-ANN exhibits a 1.37X improvement in average energy efficiency over CPU due to higher speedup than that on SIFT100M. The energy efficiency of DRIM-ANN would be further improved if dynamic gating of unused UPMEM MRAM were supported.

\subsection{Optimizations}
\label{sec:evaluation:optimizations}
\begin{figure}[!t]
\centering
\includegraphics[width=3.3in]{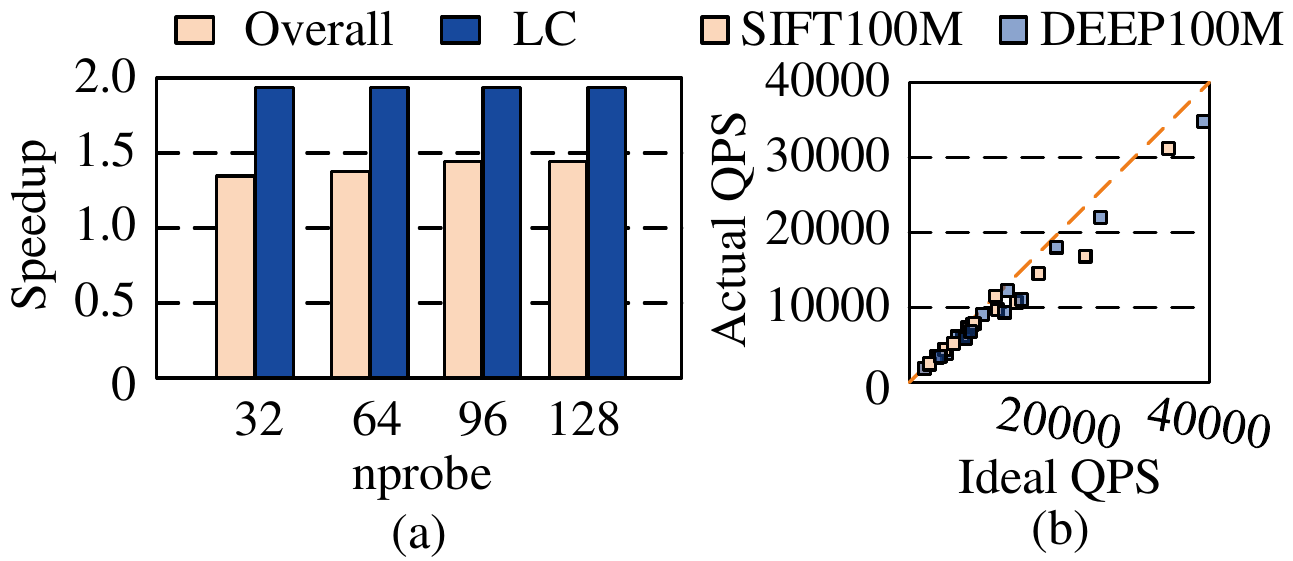}
\caption{Results of architecture-aware algorithm tuning. (a) Speedup of multiplier-less ANNS conversion. (b) Gap between actual performance and the proposed performance model.}
\label{fig14}
\end{figure}

\textbf{Multiplier-less ANNS conversion.} Figure~\ref{fig14}(a) shows the speedup with multiplier-less ANNS conversion. With the conversion, the average speedup of LC is 1.93X. Since the conversion is only applied to LC, the speedup of the end-to-end performance is 1.40X with $nlist$ set as $2^{16}$, which is slightly lower than LC. Additionally, although each primary multiplication requires 32 instructions, the speedup of LC is smaller than 32X, since the granularity of accesses to SQT is small, which hurts the real bandwidth. Despite the imperfection, the conversion still benefits from the huge bandwidth of DRAM-PIMs.

\textbf{The proposed performance model.} The performance gap between the ideal performance model and actual DRIM-ANN is shown in Figure~\ref{fig14}(b). Across all kinds of indices, the actual performance of DRIM-ANN achieves 71.80\%-99.91\% and 73.53\%-95.10\% of the predicted one on SIFT100M and DEEP100M respectively, substantiating the predictive validity of the proposed performance model. 

\begin{figure}[!t]
\centering
\includegraphics[width=3.3in]{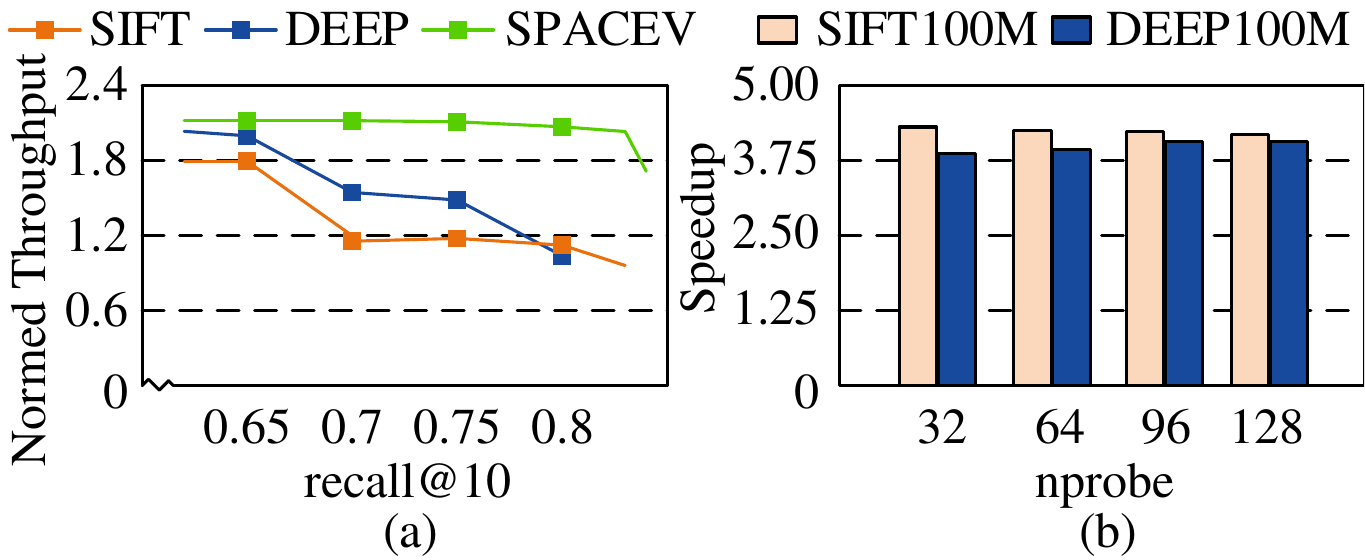}
\caption{(a) Trade-off between accuracy and performance. (b) Speedup of buffer optimization.}
\vspace{-0.2cm}
\label{fig24}
\end{figure}

\textbf{Accuracy impacts.} We evaluated DRIM-ANN’s performance under varying accuracy-constraints on SIFT100M, DEEP100M, and SPACEV100M. SPACEV100M consists of 100M 100-dimensional vectors with 29316 queries in total \cite{ref69}. The throughput results are normalized to the optimal one in Figure~\ref{fig11} where the configurations are empirically selected. Experimental results in Figure~\ref{fig24}(a) demonstrate the trade-off between accuracy and performance across all datasets.

\textbf{Buffer optimization.} We compare the performance of DRIM-ANN with/without WRAM utilization to verify the benefits of the proposed buffer optimization. As shown in Figure~\ref{fig24}(b), DRIM-ANN improves the performance by 4.18X-4.30X and 3.86X-4.07X on SIFT100M and DEEP100M respectively through buffer optimization. This is because there are many random memory accesses for ANNS such as retrieval of SQT, LUT and metadata of cluster slices, which hurt the bandwidth of DRIM-ANN without WRAM utilization. Besides, the geomean speedups are 4.24X and 3.98X on SIFT100M and DEEP100M respectively, which approach the upper bound since the peak bandwidth of WRAM buffers is just about 4.72X higher than ordinary MRAM memory, indicating the effectiveness of the proposed buffer optimization.

\begin{figure}[!t]
\centering
\includegraphics[width=3.3in]{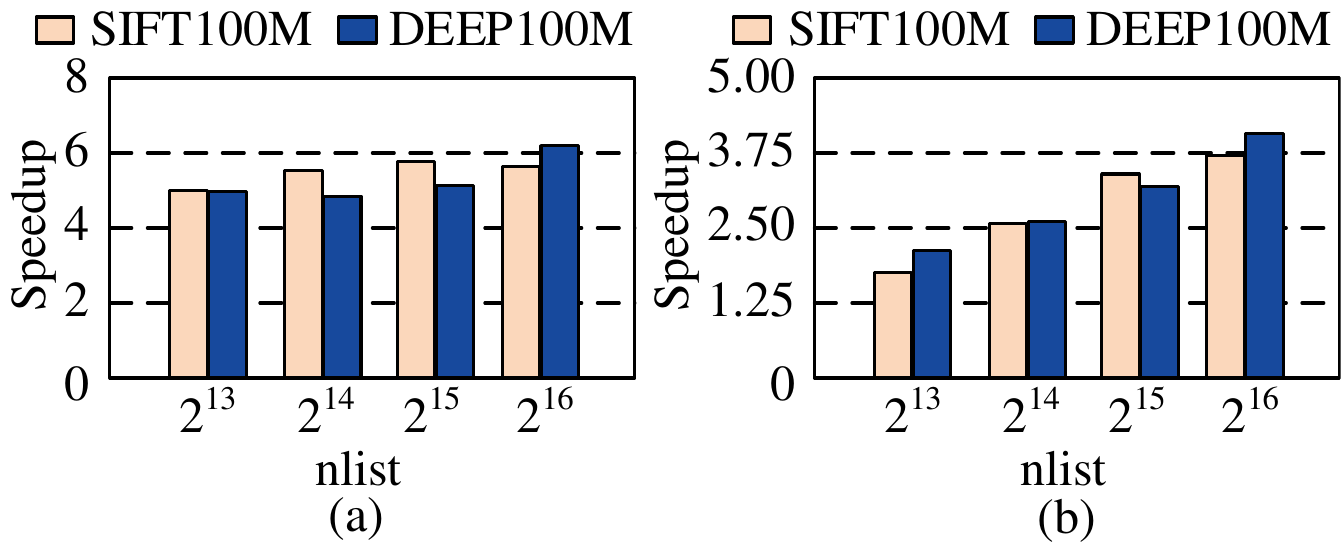}
\caption{Speedup of the proposed load-balance optimization. (a) The overall speedup. (b) Speedup of cluster allocation.}
\label{fig40}
\end{figure}

\begin{figure}[!t]
\centering
\includegraphics[width=3.3in]{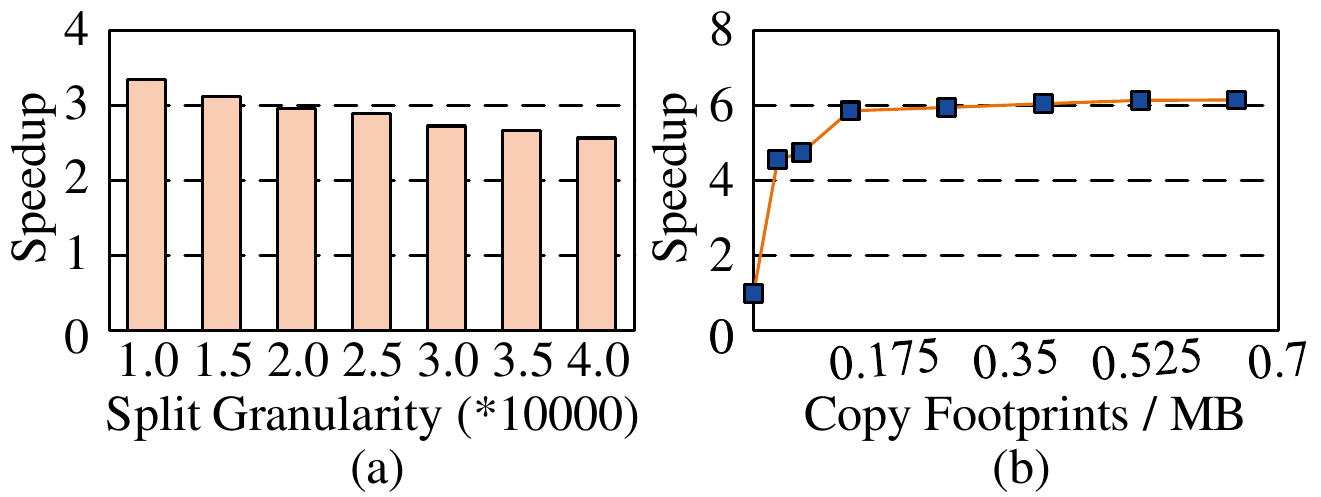}
\caption{(a) Speedup of cluster allocation and partition with various split granularity. (b) Speedup of cluster allocation and duplication with the increase of extra footprints on each DPU.}
\label{fig41}
\end{figure}

\textbf{Load-balance optimization.} Figure~\ref{fig40} and Figure~\ref{fig41} illustrate the benefits from the proposed load-balance strategies. As shown in Figure~\ref{fig40}(a), compared to the imbalanced version, the overall speedup ranges from 4.84X to 6.19X with the increase of $nlist$. Specifically, as shown in Figure~\ref{fig40}(b), the speedup with cluster allocation optimization ranges from 1.76X to 4.07X. With cluster allocation optimization applied, as shown in Figure~\ref{fig41}(a) and Figure~\ref{fig41}(b) respectively, the speedup of cluster partition achieves up to 3.35X, and the speedup of cluster duplication becomes saturated when the extra memory footprints on each DPU increase to 0.129MB which is no more than 20\% of the whole dataset. These strategies make the overall speedup of load-balance optimization shown in Figure~\ref{fig40}(a) approach the ideal speedup indicated in Figure~\ref{fig14}(b).

\subsection{Scalability}

\begin{figure}[!t]
\centering
\includegraphics[width=3.3in]{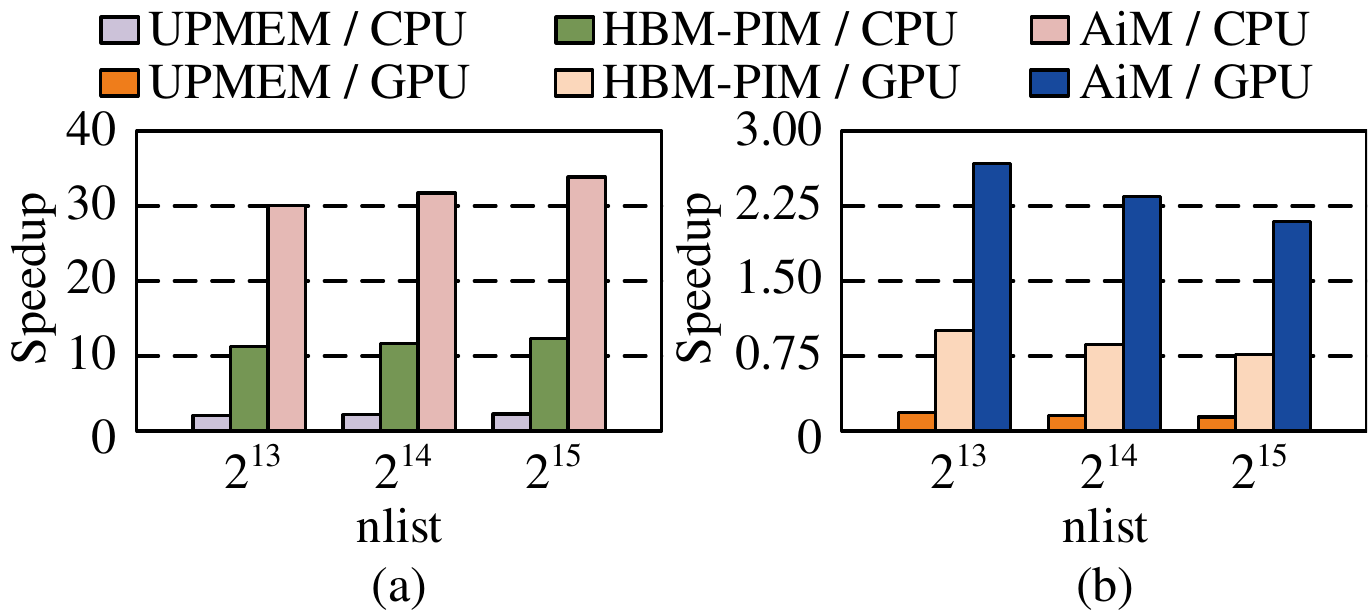}
\caption{(a) Speedup of DRIM-ANN on UPMEM, HBM-PIM and AiM over Faiss-CPU on SIFT100M. (b) Speedup of DRIM-ANN on UPMEM, HBM-PIM and AiM over Faiss-GPU on SIFT100M.}
\label{fig21}
\end{figure}

\textbf{Comparison with GPU.} Based on the indices utilized in Figure~\ref{fig11}, Faiss-GPU is about 12.33X faster than Faiss-CPU, outperforming DRIM-ANN on UPMEM, due to the rather poor computational ability of UPMEM which is only about 0.54\% of A100. Meanwhile, the peak bandwidth of A100 is more than 1.25X higher than UPMEM. Thus, the performance of DRIM-ANN on UPMEM is only about 16.00\% of Faiss-GPU. 
To verify the scalability of DRIM-ANN on other PIM products with higher computational ability, we scaled the computational ability and bandwidth from UPMEM to Samsung's HBM-PIM \cite{ref52} and SK-Hynix's AiM \cite{ref53} which only support simulation for now, although their computational abilities are only 3.69\% and 12.31\% of A100 respectively, which are still much poorer than GPU. Both of them embed computing units into HBM, which is also equipped by the baseline NVIDIA A100 GPU platform. As shown in Figure~\ref{fig21}(a), DRIM-ANN on HBM-PIM and AiM achieves 11.29X-12.30X and 30.11X-33.86X speedup over Faiss-CPU on SIFT100M respectively. The geomean speedups of DRIM-ANN on HBM-PIM and AiM are 11.74X and 31.86X respectively. As for performance comparison with GPU, as shown in Figure~\ref{fig21}(b), DRIM-ANN on HBM-PIM and AiM achieves 0.76X-1.00X and 2.09X-2.67X speedup over Faiss-GPU on SIFT100M and the geomean speedups are 0.86X and 2.35X respectively. The results indicate the effectiveness of the proposed optimization strategies for PIM products and indicate that DRIM-ANN has potential for higher performance on future DRAM-PIMs with better computational ability. 

\textbf{Comparison with other PIM-solutions.} We noticed that MemANNS \cite{ref70} is a contemporaneous work that also explores large-scale ANNS on UPMEM. Since MemANNS is not open-sourced, we use the performance data reported in MemANNS paper for comparison. As shown in table~\ref{table4}, DRIM-ANN performs slightly below MemANNS under linear scaling assumptions, but surpasses it when DSE is applied. While both approaches share common optimizations like hot cluster duplication and query scheduling, DRIM-ANN introduces additional techniques like PIM-aware algorithm DSE, SQT, and partition-based load balancing that contribute to its competitive performance.

\begin{table}[!t]
  \centering
  \caption{Comparison with MemANNS \cite{ref70} on SIFT1B.}
  \label{table4}
  \begin{tabular}{l@{\ \ \ }c@{\ \ \ }c}
    \hline
    \textbf{} & \textbf{\#DPUs} & \textbf{QPS(SIFT1B)}\\
    \hline
    MemANNS \cite{ref70} & 896 & 405 \\
    DRIM-ANN (without DSE) & 1018 & 419 \\
    DRIM-ANN (with DSE) & 1018 & \textbf{3867} \\
    \hline
  \end{tabular}
\end{table}

\textbf{Larger datasets.} We also verify the scalability of DRIM-ANN on larger datasets by expanding SIFT1B through dimensions. Our experiments demonstrate that when the datasets exceed the GPU memory and approach the limit of UPMEM memory on our server, DRIM-ANN on UPMEM still works and keeps the speedup over the CPU baseline, while the GPU baseline fails due to memory overflow since it requires the dataset to be fully loaded into GPU memory \cite{ref50}. Meanwhile, using multiple GPUs for ANNS causes performance degradation according to \cite{ref50}. With more UPMEM DIMMs plugged on the server, DRIM-ANN would support ANNS on even larger datasets with high performance, demonstrating better scalability on large datasets than GPU.

\section{Discussion}
\textbf{SIMD optimization. } As mentioned in Section~\ref{sec:intro-PIMs}, current UPMEM chips exhibit constrained computational capacity due to restricted clock frequency and absence of vector multiplication units. While DRIM-ANN enhances multiplication through LUT replacement, distance calculation remains SISD-based (single instruction single data stream), bottlenecked by high-overhead instruction processing. Emerging PIM architectures such as HBM-PIM \cite{ref52} and PIM-AI \cite{ref66} alleviate this limitation by incorporating SIMD-capable vector units (single instruction multiple data), introducing trade-offs between computational enhancement and pattern-based LUT replacement.

\textbf{Lock pruning. } DRIM-ANN implements a top-k engine on DPUs to reduce distance results requiring host-side merging. The top-k priority queue is shared across all tasklets within a DPU, necessitating lock protection. The locked top-k priority queue causes considerable locking overhead, which reaches approximately 50\% of total latency in certain scenarios. To alleviate stalls induced by comparison between distance results and top-k records, DRIM-ANN forwards the current top-k record to DC. Although not guaranteed to be the latest, the forwarded record suffices for lock pruning. As shown in Figure~\ref{fig13}, the overhead of TS becomes negligible with top-k forwarding.

\textbf{Various ANNS algorithms. } Cluster-based ANNS algorithms generally consist of two stages: cluster locating(CL) and cluster searching(CS). For example, QuickNN \cite{ref60}, SPANN \cite{ref56}, HCNN \cite{ref67}, and SmartANNS \cite{ref68} use tree- or graph-based structures for cluster organization. As illustrated in Figure~\ref{fig3}, DRIM-ANN performs IVF-based CL on host CPU and PQ-based CS on PIM-DIMMs. This proposed design allows easy adaptation to other cluster-based ANNS methods by replacing CPU-side CL while reusing the PIM-DIMM acceleration for CS in future work.

\section{Related Works}
\subsection{Approximate Nearest Neighbor Search}
To alleviate the huge computational complexity and memory accesses caused by the large amount of distance calculation of nearest neighbor search, a variety of ANNS algorithms have been proposed, including tree-based \cite{ref59, ref60, ref38, ref61}, hash-based \cite{ref62, ref63, ref64}, graph-based \cite{ref17, ref16, ref35, ref37} and cluster-based \cite{ref23, ref58, ref5, ref50, ref8} ones. For ANNS on high-dimensional vector corpus, recent works have proved that cluster-based ANNS is the most efficient on billion-scale datasets with the same memory footprint \cite{ref50, ref7, ref56}, and it is especially friendly to dynamic vector data \cite{ref1}, which is eagerly required by online applications. Among cluster-based ANNS libraries, Faiss developed by Meta is the most popular GPU accelerated one \cite{ref8} and has made efforts on ANNS optimization based on inverted file with product quantization (IVF-PQ) and its variants for many years \cite{ref9, ref4}, which was chosen as the baseline of our experiments. Besides the industrial-grade library, the academia has also proposed many works for ANNS acceleration on CPU \cite{ref19, ref23, ref24, ref25, ref56, ref57, ref58}, GPU \cite{ref5, ref50, ref27, ref39} and accelerators \cite{ref7, ref8}. Among them, \cite{ref19, ref23, ref24} and \cite{ref25} explore fast index construction or searching with pre-constructed LUT or SIMD instruction on CPU. VDTuner \cite{ref58} explores the optimal parameters of ANNS algorithms with Bayesian optimization and puts both performance and accuracy into optimizing iterations instead of estimating the performance with an explicit model due to the complexity of modeling with CPU instruction set architecture and cache optimization. On GPU, \cite{ref27} presents Product Quantization Tree (PQT) for large scale dataset, and \cite{ref39} proposes a novel generic inverted index framework. JUNO \cite{ref5} filters redundant LUT construction with dynamic distance thresholds based on the position of queries and maps the system to GPU Tensor-RT cores. RUMMY \cite{ref50} divides clusters into small slices based on a given standard deviation to balance workloads among thread blocks and reorders clusters greedily to  maximize the overlapping between PCIe transmission and GPU computation. As for designs of accelerators, ANNA \cite{ref7} proposes a fixed ASIC design for IVF-PQ supporting multiple compression ratio and verifies the design with a custom cycle-level simulator. FANNS \cite{ref8} searches for legal IVF-PQ parameters exhaustively with the user-provided accuracy constraint and a sample query set and predicts the resource consumption and throughput of the accelerator by a combination of basic hardware building blocks on one of several kinds of FPGA devices to generate the optimal design. However, development and optimization for the specific designs of accelerators are laborious, which limits the application of them. For general-purpose processors, the bandwidth of ordinary memory chips on CPU is limited, while the capacity of GPU memory is bounded, which limits the scalability of deployment on them. Compared to them, DRIM-ANN is designed for general-purpose PIM processors, which strikes a balance between the scalability and laborious development.

\subsection{Processing-in-memory Architectures}
PIM technologies are proposed to address the "Memory-Wall" problem. They are mainly categorized into two types: (1) PIMs built with die-stacking memories such as High Bandwidth Memory (HBM) and Hybrid Memory Cube (HMC), where PIM computation is completed within a logic die; (2) PIMs built with Dual-Inline
Memory Modules (DIMMs) such as UPMEM, where PIM computation is supported by DPUs inner DRAM chips. Both solutions have attracted tremendous interests in recent years \cite{ref16, ref20, ref41, ref42, ref43, ref44, ref45, ref46, ref47}.

To alleviate the memory bottleneck of ANNS, previous works have proposed several solutions on PIM platform. For instance, GCiM \cite{ref20} implements specific modules on the logic die of stacked memory and applies data layout optimization to graph construction. Spitfire \cite{ref44} proposes a linear-time partitioning algorithm for load balance on distributed PIM system and optimizes ANNS processing through query partitioning, calculation pruning and candidate set refinement. CXL-ANNS \cite{ref46} deploys ANNS in the CXL memory network and proposes a collaborative kNN search design for the system. Though processing in die-stacking memories can also attain huge bandwidth, the capacity is bounded. Although several works have tried to deploy ANNS on PIM-DIMMs with simulators \cite{ref47} or operational prototype systems \cite{ref46}, DRIM-ANN is the first ANNS framework deployed on commodity DRAM-PIMs to our best knowledge.

\section{Conclusion}
In this paper, we propose DRIM-ANN, the first cluster-based ANNS framework based on commercial off-the-shelf DRAM-PIMs. We propose a PIM-aware algorithm tuning strategy in-combination with lossless multiplier-less conversion and buffer/pipeline optimization to address the computing bottleneck of DRAM-PIMs substantially. In addition, we propose a systematic data layout optimization strategy along with a runtime scheduling algorithm to address the load imbalance challenge faced in deploying ANNS on the massively parallel DPUs in DRAM-PIMs. Compared with the state-of-the-art ANNS on multicore processors, DRIM-ANN achieves up to $2.46\times$ speedup, and demonstrates significant potential of further acceleration on future DRAM-PIM architectures.

\begin{acks}
We thank all the anonymous reviewers for their insightful feedback, and Prof.Xueqi Li and Prof.Guangyu Sun, for generously providing access to local UPMEM servers during the cloud service outage. This work is supported in part by the Strategic Priority Research Program of the Chinese Academy of Sciences under Grant No.: XDB0660000, XDB0660100, XDB0660102, and XDB0660103, and the National Key Research and Development Program of China under Grant No.: 2022YFB4500405, and the National Natural Science Foundation of China (NSFC) under Grant No.: 62202453. Cheng Liu is the corresponding author.
\end{acks}


\bibliographystyle{ACM-Reference-Format}
\bibliography{References/allTheRefs}

\end{document}